\documentclass[
    ,final            
  ]
  {aipproc}

\layoutstyle{8x11single}

\def\be{\begin{equation}}
\def\ee{\end{equation}}
\def\bea{\begin{eqnarray}}
\def\eea{\end{eqnarray}}

\begin{document}

\title{SMWDs as SGRs/AXPs and the lepton number violation}



\author{J.~Adam,~Jr.}{address={Institute of Nuclear Physics ASCR, CZ--250 68 \v{R}e\v{z}, Czechia}}
\author{V.B.~Belyaev \footnote{deceased}\ }{
  address={Bogolyubov Laboratory of Theoretical Physics, Joint
Institute for Nuclear Research, Dubna  141980, Russia} }
\author{P. Ricci}{
  address={Istituto Nazionale di Fisica Nucleare, Sezione di
Firenze, I-50019 Sesto Fiorentino (Firenze), Italy } }
\author{F.~\v{S}imkovic}{
  address={Department of Nuclear Physics and Biophysics,
Comenius University, Mlynsk\'a dolina F1, SK--842 15, Bratislava,
Slovakia}, altaddress={Bogolyubov Laboratory of Theoretical Physics,
Joint Institute for Nuclear Research, Dubna  141980, Russia} }
\author{E.~Truhl\'{\i}k}{address={Institute of Nuclear Physics ASCR, CZ--250 68 \v{R}e\v{z}, Czechia}}

\begin{abstract}
Possible nature of strongly magnetized white dwarfs (SMWDs) is
studied. It is shown that for relatively low values of the
equatorial surface magnetic field $B\,\sim\,10^9\,-\,10^{11}$ G they
can be good candidates for soft gamma-ray repeaters and anomalous
X-ray pulsars (SGRs/AXPs). For the case of iron SMWDs the influence
of a neutrinoless electron to positron conversion on the SGRs/AXPs
luminosity is estimated.
\end{abstract}

\maketitle


\section{Introduction}

It has recently been shown that SMWDs can be progenitors of the
superluminous supernovae Ia \cite{DM1,DM2,BEBH}. In parallel, we
have applied in Refs.\,\cite{AEAL1,AEAL2} the concept of the SMWDs,
developed in Ref.\,\cite{DM1}, to the study of the reaction of the
double charge exchange
\be
\mathrm{e}^- + \mathrm{X}(A,Z)
\rightarrow \mathrm{X}(A,Z-2) + \mathrm{e}^+ \,,  \label{EMEPC}
\ee
which can happen only if the neutrinos are of the Majorana type,
implying thus the lepton number violating process of electron
capture by a nucleus X($A,Z$), which was in our case $^{56}_{26}$Fe.

In Table 4 \cite{AEAL2}, we presented  the ratio of the calculated
change in the luminosity $\Delta\, L$ of the SMIWDs to the solar
luminosity $L_\odot$, employing the necessary input from Table 2
\cite{AEAL2}, $|\langle m_\nu\rangle|$=0.4 eV and 0.8 eV, and the
nuclear radius $R$=1.2~A$^{1/3}\,\approx\,$ 4.59 fm. For
convenience, we present the results for $\Delta\, L$ below, labeled
by the corresponding absolute value of the effective mass of
Majorana neutrinos $|\langle m_\nu\rangle|$.
\begin{itemize}
\item
For the Fermi energy $E_{\,\mathrm{F}}$ = 20 $m_\mathrm{e}$, \be
(\Delta L)_{0.4} = 6.07\,\times 10^{18} {\rm erg\,s}^{-1}\,,\quad
(\Delta L)_{0.8} = 2.42\,\times 10^{19} {\rm erg\,s}^{-1}\,.
\label{FE20} \ee
\item
For the Fermi energy $E_{\,\mathrm{F}}$ = 46 $m_\mathrm{e}$, \be
(\Delta L)_{0.4} = 7.64\,\times 10^{21} {\rm erg\,s}^{-1}\,,\quad
(\Delta L)_{0.8} = 3.04\,\times 10^{22} {\rm erg\,s}^{-1}\,.
\label{FE46} \ee
\end{itemize}

Here, $m_\mathrm{e}$ is the electron mass.

\section{SMIWDs as SGRs/AXPs}

During the last decade the observational astrophysics has made
substantial progress in the study of the SGRs/AXPs sources. The
McGill Magnetar Catalog \cite{MCG} contains 26 such objects called
magnetars. Generally, it is believed that SGRs/AXPs are the neutron
stars (NSs), powered by the decay of strong surface magnetic fields
of the order up to 10$^{15}$~G \cite{DT,TAD}. They are specified by
a long rotational period $P\,\sim\,(2-12)$~s and by its time
derivative \mbox{$\stackrel{.}P\,\sim\,$
(10$^{-11}$-10$^{-15})$~ss$^{-1}$}, larger than for ordinary pulsars
with $\stackrel{.}P\,\sim\,10^{-15}$~ss$^{-1}$.

Let us note that in the pulsar model, the observed X-ray
luminosity $L_\mathrm{X}$ is supposed to come from the loss of the
rotational energy of the NS
\be
\stackrel{.}{E}^{\mathrm{NS}}_{\mathrm{rot}} = -4\pi^2 I
\frac{\stackrel{.}{P}}{P^3}\,, \label{ERO}
\ee
where $I$ is the momentum of inertia of the NS. Besides, the
surface dipolar magnetic field strength at the equator,
$B_\mathrm{e}$, and the characteristic age of the pulsar, $\tau$,
are
\be
B_ \mathrm{e}= \left( \frac{3 c^3 I}{8 \pi^2}\,\frac{P \stackrel{.}{P}}{R^6}\,
\right)^{1/2}\,\equiv\,\frac{m}{R^3}\,,\quad \tau =
\frac{P}{2\stackrel{.}{P}}\,,\label{BT}
\ee
where $R$ is the radius
of the star at the equator, $c$ is the velocity of light and $m$ is
the magnetic moment of the rotating magnetized star.

In the last years, SGRs/AXPs sources were observed
\cite{MCG,RSGR,RPC,RSWIFT,SCEAL,ZEAL,RVIPT} which, if considered as
rotation\,-\,powered NSs, provides
$\stackrel{.}{E}^{\mathrm{NS}}_{\mathrm{rot}}\,<\,$$L_\mathrm{X}$
and $B_ \mathrm{e}\,\sim$ (10$^{12}$\,-\,10$^{13}$) G
$<\,B_\mathrm{c}\,=\,m^2_\mathrm{e}\,c^3/(e\,\hbar)\,=\,4.414\,\times\,10^{13}$
G. This is in discord with the magnetar model that requires inverted
inequalities. There are alternative hypotheses about the nature of
SGRs/AXPs possessing magnetic field lower than the critical field
$B_\mathrm{c}$.
E\,.g\,., it was shown in Refs.\,\cite{MAC,BIRR,CM} that these
low magnetic field magnetars can be alternatively described as
massive rapidly rotating magnetized WDs.

Analogously, we have proposed in \cite{AEAL2}  that such magnetars
can be considered to be the SMIWDs. For this study, we have chosen
two SGRs/AXPs, namely \mbox{SGR 0418+579} and \mbox{Swift
J1822.6-1606}, for which the rotational period $P$ and the spin-down
rate $\stackrel{.}{P}$ are well known
\cite{MCG,RSGR,RPC,RSWIFT,SCEAL}. Our calculations have shown that
the loss of the rotational energy of the rapidly rotating SWIMDs can
also describe the observed luminosity of these compact objects.

Here, we report on an improvement of these calculations and include
the results for the new low magnetic field compact object \mbox{3XMM
J185246.6+003317} \cite{ZEAL,RVIPT}.\\
For the pulsar model, the values of the mass and radius of the
NS are set to $M = 1.4\, M_\odot$ and $R = 10\,$~km, whereas for
the WD model, the choice of these parameters is $M = 1.4\, M_\odot$
and $R = 3000\,$~km, in accord with Refs.~\cite{MAC,CM,BIRR}. In
the approach of the SMIWDs, we take $M=2\, M_\odot$ and the radii
from our Table 3 \cite{AEAL2}.

Next we analyze the data for the above mentioned compact objects.

\begin{itemize}
\item
 \underline{SGR 0418+5729} (data from
$^a$)\,\,Ref.\,\cite{RSGR}\,,\quad $^b$)\,\,Ref.\,\cite{RPC})
\bea
 P(s)&=&9.0784\,\,^a)\,, \qquad
 \stackrel{.}{P}(s\,s^{-1}) = 4\times 10^{-15}\,\,^a)\ , \nonumber \\
 d&=&2\, {\rm kpc}\,^a)\,,\qquad
 \Delta L_{\mathrm{X}}=7.5\times 10^{-15}\,{\rm erg\, s}^{-1} {\rm
 cm}^{-2}\,^b)\,.
  \label{DSGR}
\eea
From $\Delta L_{\mathrm{X}}$  and the distance $d$  one obtains for
the luminosity and the age
\be L_{\mathrm{X}} = 3.6 \times 10^{30} {\rm erg\, s}^{-1}\,,\quad
\tau = 36\, {\rm Myr}\,. \label{LXTSG} \ee
As described above, one gets from these numbers the results:
\bea
B_{\mathrm{NS}}& = &6.4\times 10^{12}\,{\rm G}\,,\quad
|\stackrel{.}{E}_{\mathrm{rot}}^{\mathrm{NS}}| = 7.5\times 10^{28}
{\rm erg\,
s}^{-1}\,, \label{NSSGR} \\
B_{\mathrm{WD}}& = & 7.1\times 10^{7}\,{\rm G} \,, \quad
|\stackrel{.}{E}_{\mathrm{rot}}^{\mathrm{WD}}| = 6.7\times
10^{33} {\rm erg\, s}^{-1}\,, \label{WDSGR} \\
R_{\mathrm{SMIWD}}& = & 423\, {\rm km}\,, \quad B_{\mathrm{SMIWD}} =
4.3\, \times 10^9\,{\rm G}\,,\quad
|\stackrel{.}{E}_{\mathrm{rot}}^{\mathrm{SMIWD}}| = 1.91\, \times
10^{32}\,{\rm erg\, s}^{-1}\,, \label{SMWDSGR1} \\
R_{\mathrm{SMIWD}}& = & 186\, {\rm km}\,, \quad B_{\mathrm{SMIWD}} =
2.2\, \times 10^{10}\,{\rm G}\,,\quad
|\stackrel{.}{E}_{\mathrm{rot}}^{\mathrm{SMIWD}}| = 3.7\, \times
10^{31}\,{\rm erg\, s}^{-1}\,. \label{SMWDSGR2}
\eea

Comparing the results for the spin-down luminosity, presented in
Eqs.\,(\ref{NSSGR}) - (\ref{SMWDSGR2}), with the luminosity
$L_\mathrm{X}$ of Eq.\,(\ref{LXTSG}), one can see that the loss of
the rotational energy of the SMIWDs  as well as the loss of this
energy of the WD can explain $L_\mathrm{X}$, but  for the NS,
$|\stackrel{.}{E}_{\mathrm{rot}}^{\mathrm{NS}}|\,<\,L_\mathrm{X}$.

Let us note that according to Table 1 \cite{RSGR}, in the time
interval from July 2009 to August 2012  the luminosity of this star
diminished by 1150 times! If in the last 3 years the reduction in
the luminosity were the same then it would be now $L_\mathrm{X} =
3.0 \times 10^{27} {\rm erg\, s}^{-1}$, rather than $L_\mathrm{X} =
3.6 \times 10^{30} {\rm erg\, s}^{-1}$. In our opinion,
remeasurement of its \mbox{$\Delta L_\mathrm{X}$}, $P$, and
$\stackrel{.}P$ would be highly desirable.\\

\item
 \underline{Swift J1822.6-1606} (data from Ref.\,\cite{RSWIFT})
\bea
 P(s)&=&8.4377\,,\qquad
 \stackrel{.}{P}(s\,s^{-1}) = 8.3\times 10^{-14}\ , \nonumber\\
 d&=&5\, {\rm kpc}\,,\qquad
 \Delta L_\mathrm{X}=4\times 10^{-14}\,{\rm erg\, s}^{-1} {\rm cm}^{-2}\ .
\label{DSWIFT}
\eea
From $\Delta L_\mathrm{X}$  and the distance $d$ one obtains for the
luminosity and the age
\be L_\mathrm{X} = 1.2 \times 10^{32} {\rm erg\, s}^{-1}\,,\quad
\tau = 1.61\, {\rm Myr}\ . \label{LXTSW} \ee
However, as argued by Scholz et al. \cite{SCEAL}, \mbox{Swift
J1822.6-1606} could have a comparable distance to that of the
Galactic region M17, which is 1.6 $\pm$ 0.3 kpc. In that case, \be
L_\mathrm{X} = 6.4 \times 10^{31} {\rm erg\, s}^{-1}\,.
\label{LXTSC}
\ee
For the magnetic fields and the spin-down luminosities one obtains
\bea
B_{\mathrm{NS}}& = &2.8\,\times 10^{13}\,{\rm G}\,,\quad
|\stackrel{.}{E}_{\mathrm{rot}}^{\mathrm{NS}}| = 1.9\times 10^{30}
{\rm erg\,
s}^{-1}\,, \label{NSSW} \\
B_{\mathrm{WD}}& = & 3.1\times 10^{8}\,{\rm G} \,, \quad
|\stackrel{.}{E}_{\mathrm{rot}}^{\mathrm{WD}}| = 1.7\times
10^{35} {\rm erg\, s}^{-1}\,, \label{WDSW} \\
R_{\mathrm{SMIWD}}& = & 423\, {\rm km}\,, \quad B_{\mathrm{SMIWD}} =
1.9\, \times 10^{10}\,{\rm G}\,,\quad
|\stackrel{.}{E}_{\mathrm{rot}}^{\mathrm{SMIWD}}| = 5.0\, \times
10^{33}\,{\rm erg\, s}^{-1}\,, \label{SW5} \\
R_{\mathrm{SMIWD}}& = & 186\, {\rm km}\,, \quad B_{\mathrm{SMIWD}} =
9.8\, \times 10^{9}\,{\rm G}\,,\quad
|\stackrel{.}{E}_{\mathrm{rot}}^{\mathrm{SMIWD}}| = 9.6\, \times
10^{32}\,{\rm erg\, s}^{-1}\,. \label{SW6}
\eea
Comparing the results for the spin-down luminosity, presented in
Eqs.\,(\ref{NSSW}) - (\ref{SW6}), with the luminosity $L_\mathrm{X}$
of Eq.\,(\ref{LXTSC}) one can see that the loss of the rotational
energy of the SMIWDs  as well as the loss of this energy of the WD
can explain $L_\mathrm{X}$, but  for the NS,
$|\stackrel{.}{E}_{\mathrm{rot}}^{\mathrm{NS}}|\,<\,L_\mathrm{X}$.

\item
\underline{3XMM J185246.6+003317} \\
This low magnetic field magnetar was discovered first by Zhou et al.
(2014 - \cite{ZEAL}). Its phase-coherent  timing analysis was later
redone by Rea et al. (2014 - \cite{RVIPT}). The results of both
works are similar. We restrict ourselves with the data from
Ref.\,\cite{RVIPT}:
\bea
 P(s)&=&11.5587\,,\qquad
 \stackrel{.}{P}(s\,s^{-1}) < 1.4\times 10^{-13}\ , \nonumber \\
 d&=&7.1\, {\rm kpc}\,,\qquad
 \Delta L_\mathrm{X} < 6.64\times 10^{-13}\,{\rm erg\, s}^{-1} {\rm cm}^{-2}\, .
 \label{D3XMM}
\eea
Let us note that the distance $d=7.1$ kpc in \cite{RVIPT}  was
adopted from \cite{ZEAL}.
From $\Delta L_\mathrm{X}$  and the distance $d$  of
Eq.\,(\ref{D3XMM}) one obtains for the luminosity and the age
\be
L_\mathrm{X} < 4.0 \times 10^{33} {\rm erg\, s}^{-1}\,,\quad
\tau > 1.31\, {\rm Myr}\, . \label{LX3XMM}
\ee
However, as argued by Rea  et al. \cite{RVIPT}, suggested distance
of 7.1 kpc for 3XMM J185246.6+003317 from the similarity of its
foreground absorption and the near supernova remnant Kesteven 79 can
be misleading, since the supernova remnant is much younger. Then for
\be
d  =  5\,{\rm kpc}\quad  L_\mathrm{X} < 2.0 \times 10^{33} {\rm
erg\, s}^{-1}\,, \label{3XMM5}
\ee
and for
\be
d=2\,{\rm kpc}\quad L_\mathrm{X} < 3.2 \times 10^{32} {\rm erg\,
s}^{-1}\,.   \label{3XMM2}
\ee
For the magnetic field and the spin-down luminosity one obtains
\bea
B_{\mathrm{NS}}& < &4.3\,\times 10^{13}\,{\rm G}\,,\quad
|\stackrel{.}{E}_{\mathrm{rot}}^{\mathrm{NS}}| < 1.3\times 10^{30}
{\rm erg\,
s}^{-1}\,, \label{3XMNS} \\
B_{\mathrm{WD}}& < & 4.8\times 10^{8}\,{\rm G} \,, \quad
|\stackrel{.}{E}_{\mathrm{rot}}^{\mathrm{WD}}| < 1.1\times
10^{35} {\rm erg\, s}^{-1}\,, \label{3XMWD} \\
R_{\mathrm{SMIWD}}& = & 423\, {\rm km}\,, \quad B_{\mathrm{SMIWD}} <
2.9\, \times 10^{10}\,{\rm G}\,,\quad
|\stackrel{.}{E}_{\mathrm{rot}}^{\mathrm{SMIWD}}| < 3.2\, \times
10^{33}\,{\rm erg\, s}^{-1}\,, \label{3XMSM1} \\
R_{\mathrm{SMIWD}}& = & 186\, {\rm km}\,, \quad B_{\mathrm{SMIWD}} <
1.5\, \times 10^{11}\,{\rm G}\,,\quad
|\stackrel{.}{E}_{\mathrm{rot}}^{\mathrm{SMIWD}}| < 6.3\, \times
10^{32}\,{\rm erg\, s}^{-1}\,. \label{3XMSM2}
\eea
Comparing the results for the spin-down luminosity with the
$L_\mathrm{X}$ one can see that the loss of the rotational energy of
the SMIWDs can explain $L_\mathrm{X}$ for $R_{\mathrm{SMIWD}} =
423\, {\rm km}$ and $d\,\le\,5\,$kpc and for $R_{\mathrm{SMIWD}} =
186\, {\rm km}$ and $d\,\le\,2\,$kpc, whereas the WD model can
explain $L_\mathrm{X}$ also for $d\,=\,7.1\,$kpc. On the contrary,
the loss of the rotational energy of the NS is by about two orders
of the magnitude smaller.
\end{itemize}

As for a possible role of the double charge exchange reaction
(\ref{EMEPC}), comparison of the obtained spin-down luminosities
with ($\Delta L)_{0.4}$ and ($\Delta L)_{0.8}$ of Eqs.\,(\ref{FE20})
and (\ref{FE46}), respectively, shows that the energy produced by
this reaction  cannot influence sizeably the luminosity of the
compact objects, considered above as rapidly rotating SMIWDs.

\section{Discussion of the results and conclusions}

We explored  the SMIWDs as rapidly rotating stars that can be
considered as GSRs/AXPs. We have shown  that using the observational
data for the compact objects \mbox{SGR 0418+579}, \mbox{Swift
J1822.6-1606} and \mbox{3XMM J185246.6+003317}, the loss of the
spin-down luminosity calculated in the simple SWIMD model can
reproduce the observed X-ray luminosities. However, the energy
produced by the reaction of the double charge exchange (\ref{EMEPC})
cannot influence sizeably the luminosities of the compact objects
considered as rapidly rotating SMIWDs. It means that the study of
the reaction (\ref{EMEPC}) in the SMIWDs, using simple model
\cite{DM1} with the ground Landau level and at the present level of
accuracy of measurements of the luminosity and energy of the cosmic
gamma-rays cannot provide conclusive information on the Majorana
nature of the neutrino, if the absolute value of its effective mass
is \mbox{$|\langle m_\nu \rangle|$ $\le$ 0.8 eV.}

On the other hand, more realistic models of SMWDs \cite{DM2,BEBH,SM}
already exist that can be used to improve the estimate of the yield
of the reaction (\ref{EMEPC}). Besides, new observational facilities
are expected to provide soon  the data also for fainter compact
objects, in which the effect of this reaction could be observed. Let
us mentioned one of them, the satellite Gaia (in operation from
2013), whose results will have tremendous influence on many topics
in the WD research \cite{SJ}.

\begin{theacknowledgments}
This work was supported  by the Votruba-Blokhintsev Program for
Theoretical Physics of the Committee for Cooperation of the Czech
Republic with JINR, Dubna.
 F. \v S. acknowledges the support by
the VEGA Grant agency
of the Slovak Republic under the contract No. 1/0876/12.

\end{theacknowledgments}




\bibliographystyle{aipproc}   

\bibliography{bibliography}

\begin{thebibliography}{10}
\expandafter\ifx\csname natexlab\endcsname\relax\def\natexlab#1{#1}\fi
\providecommand{\enquote}[1]{``#1''}
\expandafter\ifx\csname url\endcsname\relax
  \def\url#1{\texttt{#1}}\fi
\expandafter\ifx\csname urlprefix\endcsname\relax\def\urlprefix{URL }\fi
\providecommand{\eprint}[2][]{\url{#2}}


\bibitem{DM1} U.~Das, B.~ Mukhopadhyay, \emph{Phys. Rev.} \textbf{D86}, 042001 (2012);\\
 U.~Das, B.~ Mukhopadhyay, \emph{Phys. Rev. Lett.} \textbf{110}, 071102
 (2013).
\bibitem{DM2} U.~Das, B.~ Mukhopadhyay, \emph{J. Cosmol. Astropart. Phys.}
\textbf{06}, 050 (2014);\\
U.~Das, B.~Mukhopadhyay, \emph{J. Cosmol. Astropart. Phys.}
\textbf{05}, 016 (2015).
%
\bibitem{BEBH} P.~Bera, D.~Bhattacharya, \emph{Mon. Not. R. Astron. Soc.} \textbf{445}, 3951 {2014}.
%
\bibitem{AEAL1} V.B.Belyaev, P.Ricci, F.{\v S}imkovic, J.Adam,  M.Tater, E.Truhlik,
\emph{AIP Conf. Proc.} \textbf{ 1572}, 106 (2013).
%
\bibitem{AEAL2} V.B.Belyaev, P.Ricci, F.{\v S}imkovic, J.Adam,  M.Tater, E.Truhlik,
\emph{Nucl. Phys. A} \textbf{937}, 17 (2015).
%
\bibitem{MCG} S.A.~Olausen, V.M.~Kaspi, \emph{Astrophys. J. Suppl.} \textbf{212}, 6 (2014).
%
\bibitem{DT} R.C.~Duncan, C.~Thompson, \emph{Astrophys. J. Lett.} \textbf{392}, L9 (1992).
%
\bibitem{TAD} C.~Thompson, R.C.Duncan, \emph{Mon. Not. R. Astron. Soc} \textbf{275}, 255 (1995).
%
\bibitem{RSGR} N.~Rea et al., \emph{Astrophys. J.} \textbf{770}, 65 (2013).
%
\bibitem{RPC} N.~Rea, personal communication, 2013.
%
\bibitem{RSWIFT} N.~Rea et al., \emph{Astroph. J.} \textbf{754}, 27 (2012).
%
\bibitem{SCEAL} P. Scholz et al, \emph{Astroph.  J.} \textbf{761}, 66 (2012).
%
\bibitem{ZEAL} P.~Zhou et al., \emph{Astrophys. J. Lett.} \textbf{781}, L16 (2014).
%
\bibitem{RVIPT} N.~Rea et al., \emph{Astrophys. J. Lett.}  \textbf{781}, L17 (2014).
%
\bibitem{MAC} M.~Malheiro, J.G.~Coelho, Describing SGRs/AXPs as fast
and magnetized white dwarfs, arXiv: 1307.5074.
%
\bibitem{BIRR} K.~Boshkayev, L.~Izzo, J.A.~Rueda, R.~Ruffini, 
\emph{ Astron. Astrophys.} \textbf{555}, A151 (2013).
%
\bibitem{CM} J.G.~Coelho, M.~Malheiro, \emph{Publ. Astron. Soc. Jpn.} \textbf{66}, 1 (2014).
%
\bibitem{SM} S.~Subramanian, B.~Mukopadhyay, GRMHD formulation of highly super-Chandrasekhar
rotationg magnetised white dwarfs: Stable configurations of non-spherical white dwarfs,
arXiv: 1507.01606.
%
\bibitem{SJ} S.~Jordan, \emph{ASP Conf. Series} \textbf{493}, 443 (2015).
%


\end{thebibliography}

\IfFileExists{\jobname.bbl}{}
 {\typeout{}
  \typeout{******************************************}
  \typeout{** Please run "bibtex \jobname" to optain}
  \typeout{** the bibliography and then re-run LaTeX}
  \typeout{** twice to fix the references!}
  \typeout{******************************************}
  \typeout{}
 }

\end{document}